\documentclass{pasj00}
\draft

\begin{document}
\SetRunningHead{S. Inoue et al.}{Discovery of an Overlapping Cluster in Abell 1674 Field with Suzaku}
\Received{2014/02/13}%{yyyy/mm/dd}
\Accepted{2014/05/01}%{yyyy/mm/dd}
%\Published{}%{yyyy/mm/dd}

\title{Discovery of an Overlapping Cluster in Abell 1674 Field with Suzaku}

\author{Shota~\textsc{Inoue},\altaffilmark{1}
        Kiyoshi~\textsc{Hayashida},\altaffilmark{1}
        Hiroki~\textsc{Akamatsu},\altaffilmark{2}
        Shutaro~\textsc{Ueda},\altaffilmark{1}
        Ryo~\textsc{Nagino},\altaffilmark{1}
        Hiroshi~\textsc{Tsunemi},\altaffilmark{1}
        Noriaki~\textsc{Tawa},\altaffilmark{1}%\thanks{Present address:
        %NEC Corporation, Kawasaki, Kanagawa 211-8666, Japan}
        and
        Katsuji~\textsc{Koyama}\altaffilmark{1,3}
        }
\altaffiltext{1}{%
   Department of Earth and Space Science, Graduate School of Science,
        Osaka University,\\
1-1 Machikaneyama-cho, Toyonaka, Osaka 560-0043}
\altaffiltext{2}{%
   SRON Netherlands Institute for Space Research, Sorbonnelaan 2, 3584
        CA Utrecht, The Netherlands}
\altaffiltext{3}{%
   Department of Physics, Graduate School of Science, Kyoto
        University,\\
Kitashirakawa Oiwake-cho, Sakyo-ku, Kyoto 606-8502}
\email{shota@ess.sci.osaka-u.ac.jp}

\KeyWords{galaxies: clusters: individual: Abell 1674 --- galaxies:
clusters: intracluster medium --- X-rays: galaxies: clusters --- X-rays:
individuals: Abell 1674} %Do NOT move this preamble from here!

\maketitle

\begin{abstract}
We present the results of a Suzaku observation of Abell 1674, an
 optically very rich (richness class 3) cluster cataloged as
 $z=0.1066$. We discover the He-like Fe K-shell line from the central
 region for the first time, and find that the X-ray
 spectrum  yields a high redshift of
 $0.215^{+0.007}_{-0.006}$.
 On the other hand, the spectrum of the
 southwestern region is fitted with a redshift of $0.11\pm0.02$ by
 the He-like Fe L-shell lines, consistent with the optically determined
 value. 
The gas temperature, metal abundance, and core radius of the X-ray emission in the central
region are $3.8\pm0.2$\,keV, $0.20\pm0.05$\,$Z_\odot$ and $450\pm40$\,kpc,
respectively, while those in the southwestern region are
 $2.0\pm0.2$\,keV, $0.41^{+0.17}_{-0.13}$\,$Z_\odot$ and $220^{+90}_{-70}$\,kpc, 
respectively. These parameters are typical for clusters. 
We thus conclude that Abell 1674 consists of two
 independent clusters, A1674-C at $z\sim0.22$ and A1674-SW at $z\sim0.11$, 
overlapping along the line
 of sight. The X-ray luminosities of  A1674-C within $r=2$\,Mpc 
 is  $15.9\pm0.6\,\times 10^{43}\,{\rm erg\,s^{-1}}$ in the
 $0.1-2.4$\,keV energy band, while that for A1674-SW is 
 $1.25\pm0.07\,\times 10^{43}\,{\rm erg\,s^{-1}}$. Both are consistent with 
those expected from the $L-T$ relation of clusters within a factor of 2.
This is another support for our conclusion.
\end{abstract}

\newpage
\section{Introduction}

Abell 1674 (hereafter A1674) was listed in the cluster catalogs by \citet{Abell_1958} and \citet{Abell_1989} as a richness class 3 cluster. The redshift of A1674 was determined to be 0.1055 by \citet{Schneider_1983} from the redshift of 1 member galaxy, and is cited in the ACO catalog (\cite{Abell_1989}). Struble and Rood (1999) reported the redshift to be $z=0.1066$ based on the measurement of 2 member galaxies by \citet{Huchra_1990}.

The X-ray luminosity measured with ROSAT is $L=2.55\times10^{43}\,{\rm
erg\,s^{-1}}$ within $r=1.1$\,Mpc in the $0.5-2.5$\,keV band
(\cite{Briel_1993}). In this paper, we use the Hubble constant
$H_0=70\,{\rm km\,s^{-1}\,Mpc^{-1}}$, $\Omega_m=0.2$ and
$\Omega_\Lambda=0.8$. The metal abundance is reported with the ASCA and
XMM-Newton observations (\cite{Hashimotodani_2000} and Katayama,
Hayashida and Nishino 2005),  and is less than $0.2\,Z_\odot$. These two
parameters are significantly lower than the typical values of clusters
of galaxies of this richness class (e.g., \cite{Briel_1993},
\cite{Leccardi_2008}). They thus suggested  that metals have not yet
been supplied to the hot intracluster medium (ICM), and speculated that
A1674 is at an early evolutional stage (\cite{Hashimotodani_2000};
Katayama, Hayashida and Nishino 2005). We report the result of the Suzaku satellite
(\cite{Mitsuda_2007}) observation using the X-ray Imaging Spectrometer (XIS,
\cite{Koyama_2007}) to investigate these unusual X-ray properties of the
ICM in A1674.

\section{Observation and Data Reduction}
\label{sec:DataReduction}

\begin{figure}
 \begin{center}
  \FigureFile(80mm,50mm){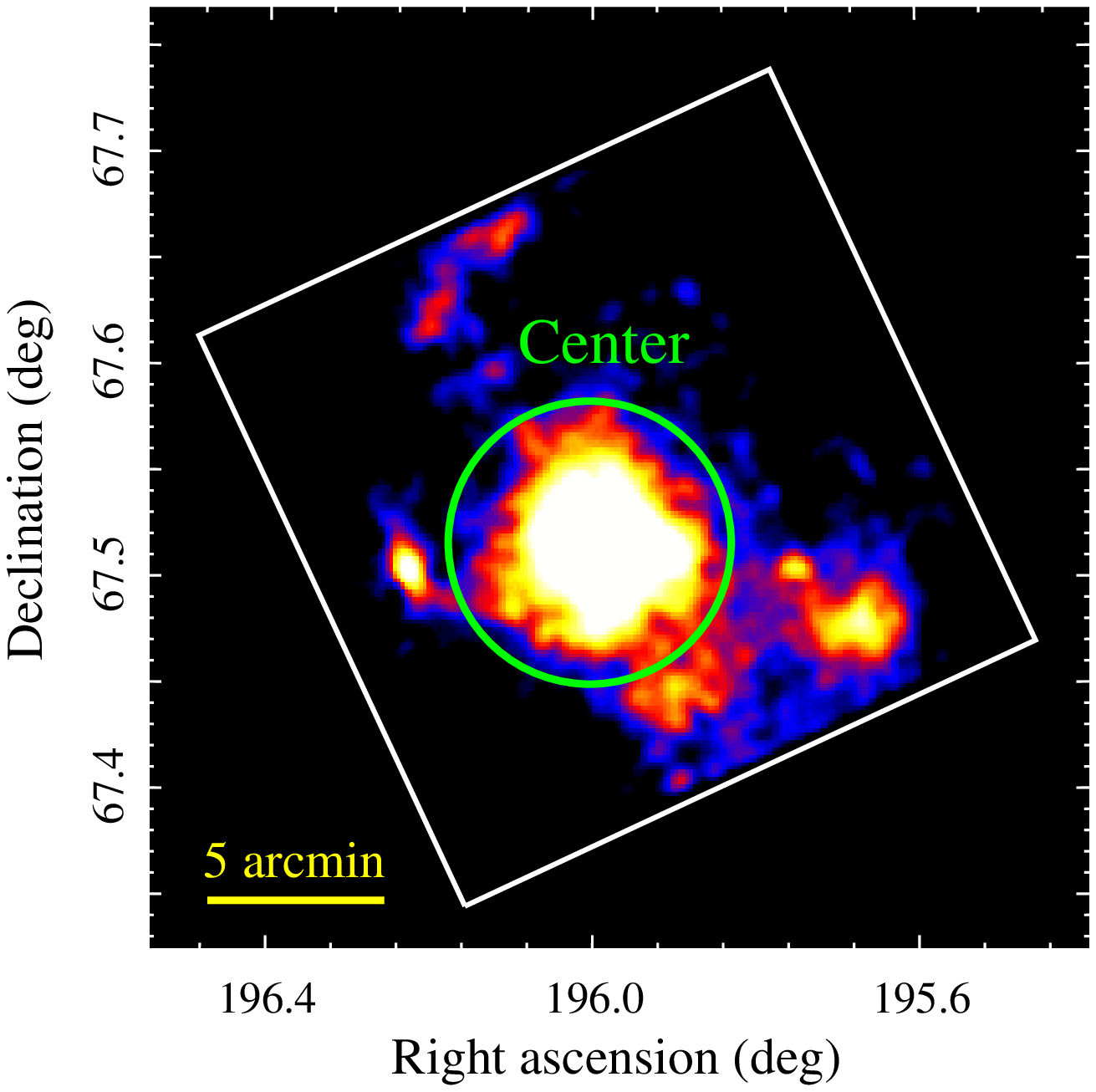} 
 \end{center}
 \caption{X-ray image of A1674, the sum of the XIS~0, XIS~1 and XIS~3 data. 
The energy ranges of the data of the FI CCDs are $0.4-10$\,keV, while
 that of the BI CCD is $0.25-8$\,keV. The green circle indicates the Center region
 defined in Sec.~\ref{subsec:Center}. The yellow solid line and
 the white box represent a 5\,arcmin scale and the Suzaku field of view, respectively.}
 \label{A1674_with_region}
\end{figure}

The observation was
performed for an effective exposure of 68 ks on 2006
December 16. Suzaku has four XIS Cameras; three of the XIS (XIS~0, 2, 3) employ front-illuminated
(FI) CCDs, and a back-illuminated CCD is installed in the other
(XIS~1). Since XIS~2 has been 
out of function from 2006 November 9, we use the remaining three XIS (XIS~0,
1, 3). The observation was performed with either the normal $3\times 3$ or
$5\times 5$ mode. These data are combined. 
Figure~\ref{A1674_with_region} shows the X-ray image of A1674.
In addition to the bright central emission, faint point-like sources in the east, and diffuse emissions in the northeast and southwest are observed.

\section{Analysis and Results}
\label{sec:results}

We extracted the X-ray spectra using XSELECT ver.2.4. 
The non-X-ray background (NXB) was produced with the FTOOL {\it{xisnxbgen}}
(\cite{Tawa_2008}), using the night-earth data taken before and after 150 days of our observation. The NXB was subtracted from the source spectra, while the X-ray background (XB) was included in the spectral fitting.
The response matrix files were generated by the FTOOL {\it{xisrmfgen}}. The auxiliary response files (ARFs) were generated by {\it{xissimarfgen}} with the observed image of A1674 in the $0.5-2$\,keV band (\cite{Ishisaki_2007}).
We analyzed the spectra in the $0.4-8.0$\,keV and the
$0.25-7.0$\,keV bands for the FI CCDs and the BI CCD, respectively.

\subsection{Spectral Analysis of the Central Region}
\label{subsec:Center}

We extracted the X-ray spectrum of the central region of
the radius of $\timeform{4'}$ with the center position at $(\alpha,
\delta)=(\timeform{196D.0035}, \timeform{67D.5164})$. The spectra
after the NXB subtraction are shown in Fig.~2. The spectral model includes the
X-ray background (XB) in addition to the ICM. The XB consists of the cosmic X-ray
background (CXB), the Milky Way halo (MWH) and the local hot bubble (LHB).
The spectrum of the CXB is assumed to be a {\it{powerlaw}} model with
the parameters presented by \citet{Kushino_2002}. The emissions from the
MWH and LHB are modeled with a XSPEC model {\it{apec}} (\cite{Smith_2001})
with metal abundances of $1$\,$Z_\odot$. We report the metal abundances
relative to the solar values in \citet{Anders_Grevesse_1989}.

We apply the {\it{apec}} model for the ICM spectra.
The absorption is given by the {\it{wabs}} model (\cite{Morrison_McCammon_1983}), where the hydrogen column density is fixed to $N_{\rm H}=1.7\times10^{20}$\,${\rm cm^{-2}}$ (\cite{Kalberla_2005}). 
All together, our spectral model is $[{\it{apec}}({\rm ICM})+{\it{powerlaw}}({\rm
CXB})+{\it{apec}}({\rm MWH})]*{\it{wabs}}+{\it{apec}}({\rm LHB})$. 

\begin{figure}
   \begin{minipage}{0.5\hsize}
   \begin{center}
      \FigureFile(80mm,50mm){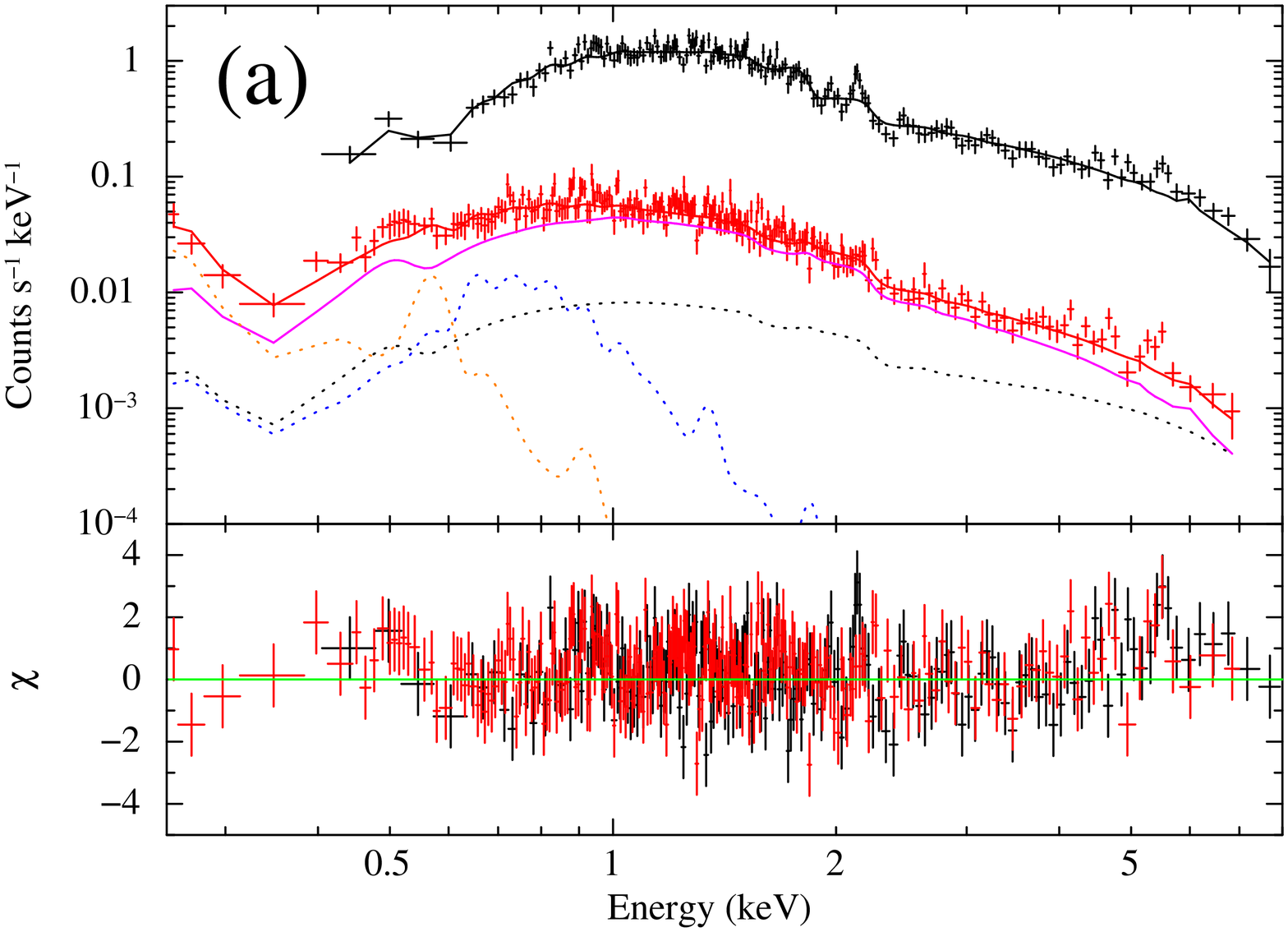}
   \end{center}
   \end{minipage}
      \begin{minipage}{0.5\hsize}
   \begin{center}
      \FigureFile(80mm,50mm){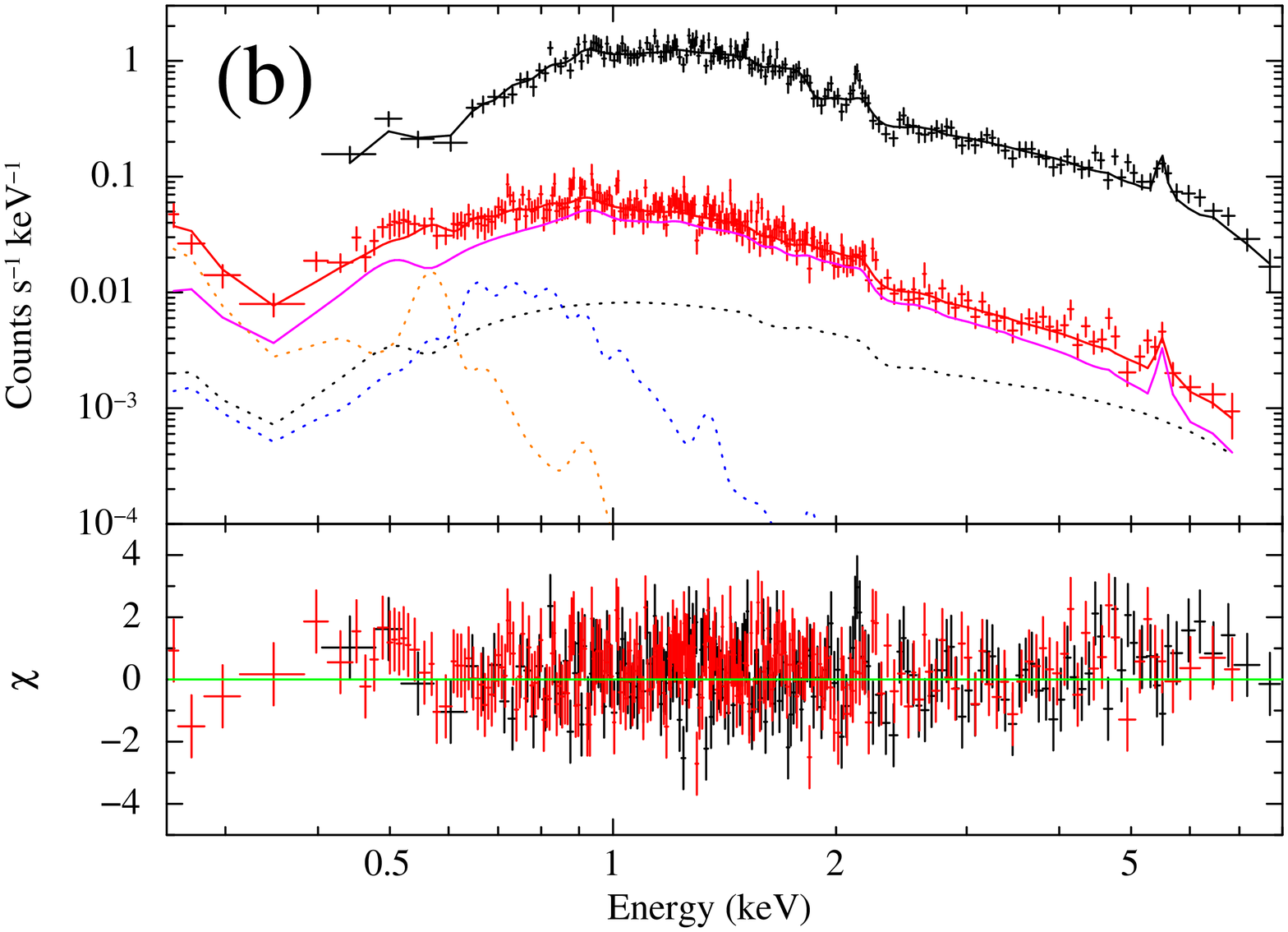}
   \end{center}
   \end{minipage}
 \caption{Spectra of the Center region after the subtraction of the NXB. The FI (black) and BI (red) spectra are fitted simultaneously with the model of ICM (magenta solid line) +  CXB (black dashed line)+ MWH (blue dashed line)+ LHB  (orange dashed line). The FI spectrum is multiplied by 30 for display purpose. In (a), we perform a model fit
with fixing $z=0.1066$, while (b) is redshift of the ICM of a free parameter.}
 \label{spec_Center}
\end{figure}

In the fitting, we first fixed the redshift to be $z=0.1066$ as
suggested by 
the optical data
(\cite{Struble_Rood_1999}). Figure~\ref{spec_Center}~(a) shows the
best-fit models. The best-fit
temperature and metal
abundances of the ICM are
 $3.6^{+0.2}_{-0.3}$\,keV and $0.04\pm0.04\,Z_\odot$,
respectively. 
These  are consistent with the previous observations
(\cite{Hashimotodani_2000}; Katayama,
Hayashida and Nishino 2005). However in Fig.~\ref{spec_Center}~(a),  we see a clear line-like residual at around $5.5$\,keV, possibly redshifted iron K-shell line. 
We then allowed  the redshift to be a free parameter.
The best-fit model is shown in Fig.~\ref{spec_Center}~(b), while the best-fit parameters are summarized in Table~\ref{tab:Center_result}.
This fit well reproduces the line-like structure at around 5.5\,keV, as
the redshifted ($z=0.215^{+0.007}_{-0.006}$) iron K-shell line. Using
the F-test, decrease in $\chi^2$ is highly significant with a null
hypothesis probability of $4\times10^{-13}$.
This is the first detection of the iron K-shell line from A1674.
The metal abundance of the Center region is $0.20\pm0.05$\,$Z_\odot$, which
is not unusual for clusters.  Obviously, the statistics of the previous
X-ray observations (\cite{Hashimotodani_2000}; Katayama,
Hayashida and Nishino 2005)
were too poor to determine the reliable redshift and metal abundance.

\subsection{Spectral Analysis of the Southwestern Region}
\begin{figure}
   \begin{minipage}{0.5\hsize}
   \begin{center}
      \FigureFile(70mm,50mm){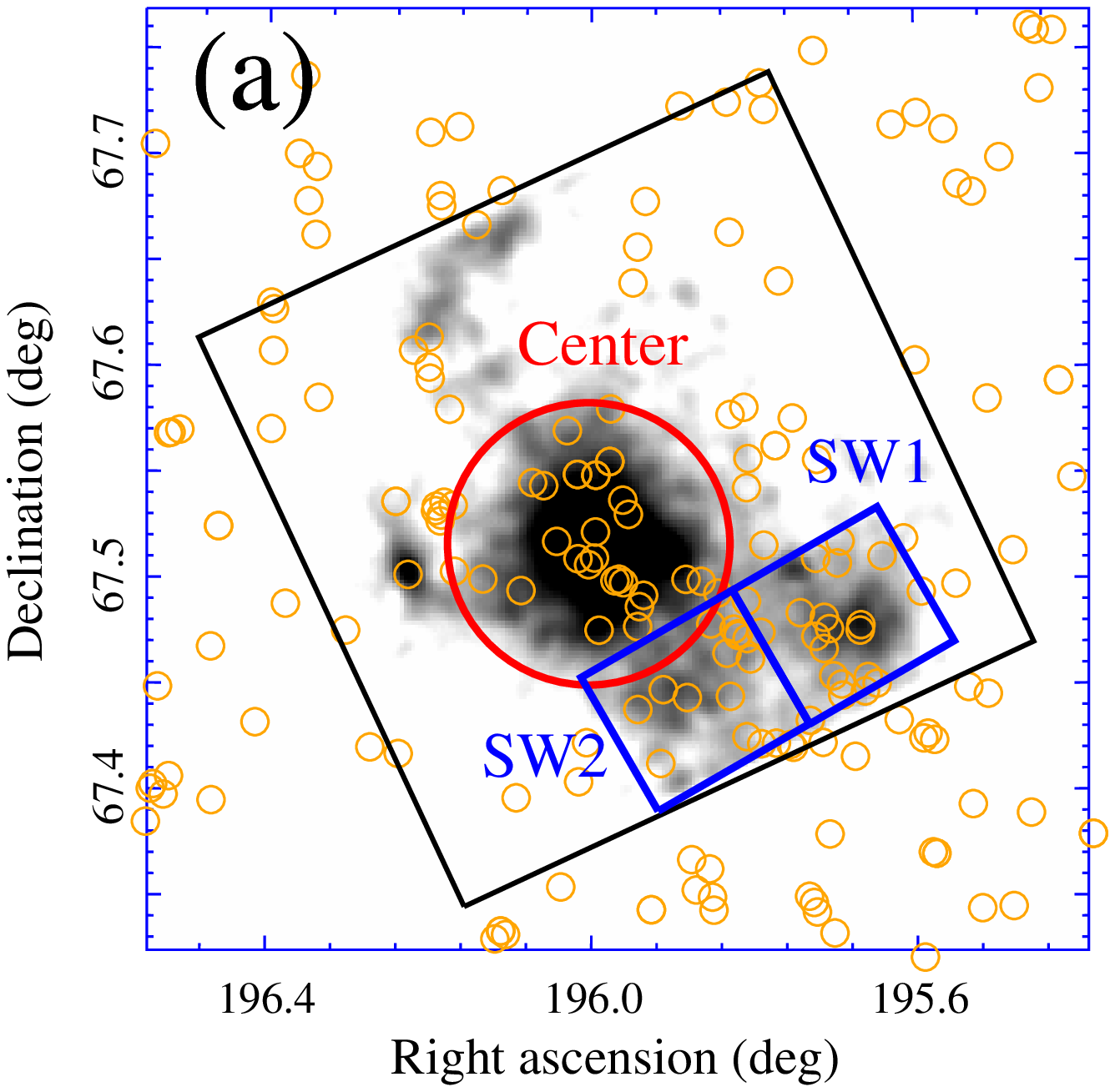}
   \end{center}
   \end{minipage}
      \begin{minipage}{0.5\hsize}
   \begin{center}
      \FigureFile(70mm,50mm){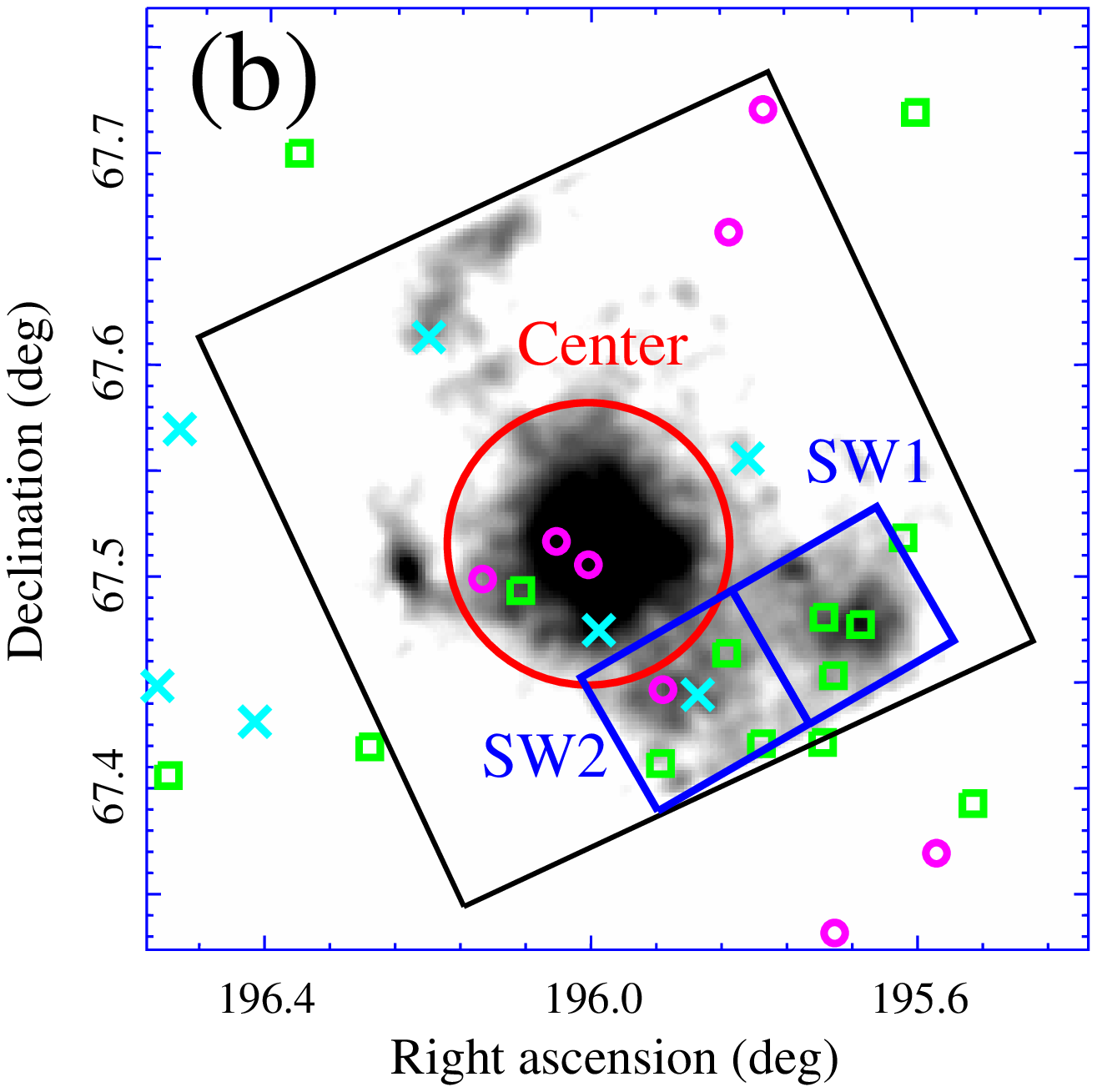}
   \end{center}
   \end{minipage}
 \caption{Galaxies observed in the SDSS plotted overlaid on the X-ray image of A1674. The red circle, two blue boxes and black box represent the Center, SW1, SW2 region and the Suzaku field of view, respectively. (a): The orange circles indicate galaxies listed in the SDSS photometry catalog, and are selected under the condition that all the magnitudes of the $g$, $r$, $i$, and $z$ bands are brighter than 20. (b): Same as (a), but is redshift-sorted galaxy distribution. The green squares and magenta circles are galaxies with $0.10<z<0.12$ and those with $0.21<z<0.23$, respectively. The galaxies with the other redshifts are indicated by cyan crosses.}
 \label{A1674_with_SDSS}
\end{figure}

Since the X-ray redshift of the  A1674 Center region
($z=0.215^{+0.007}_{-0.006}$) is different from that determined
optically, we examine the spatial distribution of optical galaxies near
and around A1674. Figure~\ref{A1674_with_SDSS}~(a) shows the galaxy
distribution  on the Suzaku X-ray image of A1674  taken from the Sloan
Digital Sky Survey (SDSS, \cite{York_2000}) database, while
Fig.~\ref{A1674_with_SDSS}~(b) is the redshift-sorted galaxy
distribution made from the SDSS spectroscopic survey list.  Although the
SDSS spectroscopic survey is not deep enough to include most of the
galaxies with $z\sim 0.2$ in this region (\cite{Strauss_2002}), we see
three galaxies with $0.21<z<0.23$ in the Center region. On the other hand, eight
galaxies with $0.10<z<0.12$ are found in the southwestern region.  
We find two separated structures in the X-ray emission in the southwestern region. 
We thus define two extraction regions, SW1 and SW2, marked by the solid blue
line in Fig.~\ref{A1674_with_SDSS}. The center coordinate of
SW1 and that in SW2 are $(\alpha, \delta)=(\timeform{195D.6696},
\timeform{67D.4784})$ and $(\timeform{195D.8993}, \timeform{67D.4470})$,
respectively. The X-ray emissions in SW1 and SW2 are more extended 
than the point-spread function (PSF) of the X-ray telescope on board Suzaku
(\cite{Serlemitsos_2007}).

\begin{figure}
   \begin{minipage}{0.5\hsize}
   \begin{center}
      \FigureFile(80mm,50mm){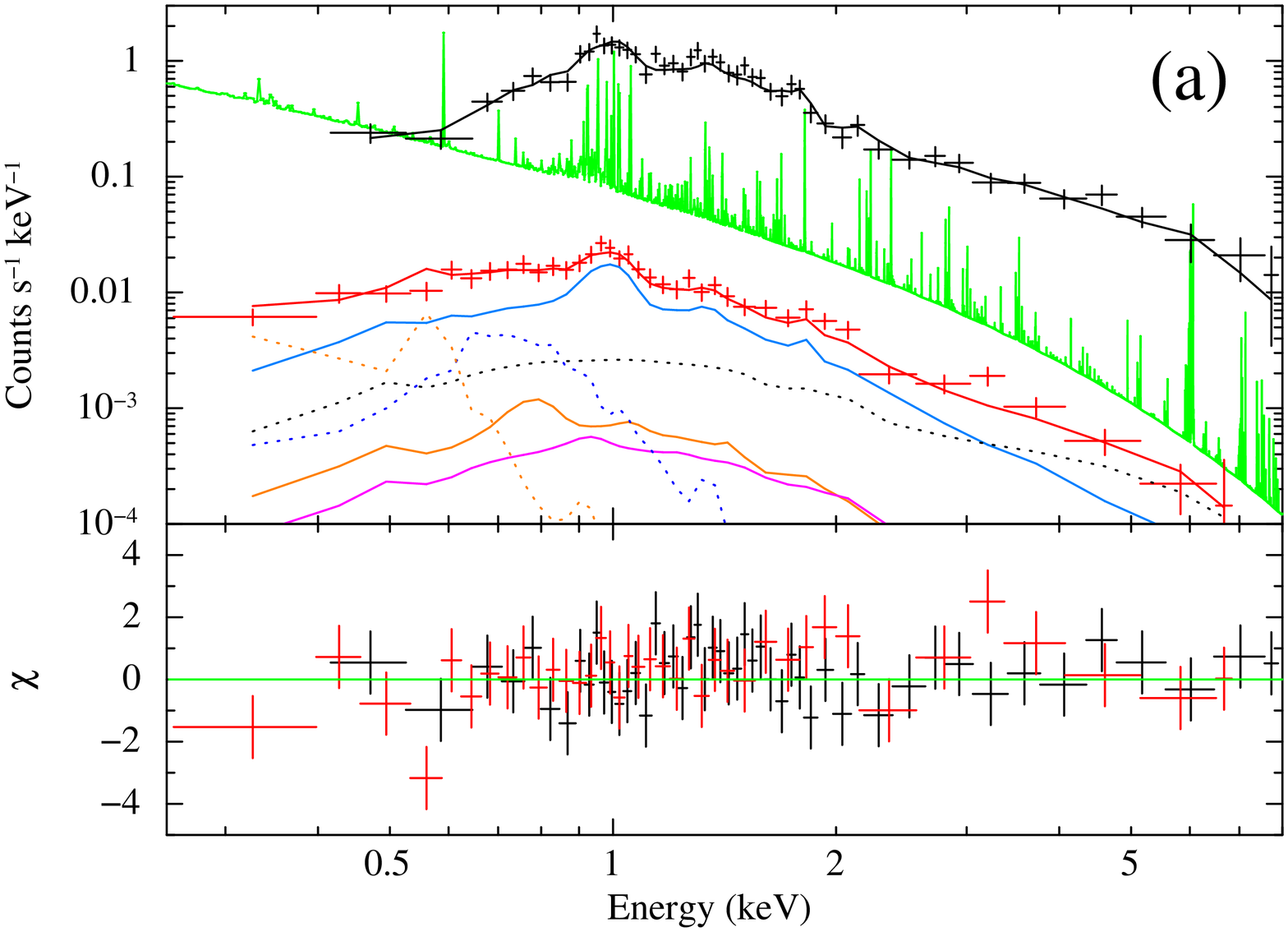}
   \end{center}
   \end{minipage}
      \begin{minipage}{0.5\hsize}
   \begin{center}
      \FigureFile(80mm,50mm){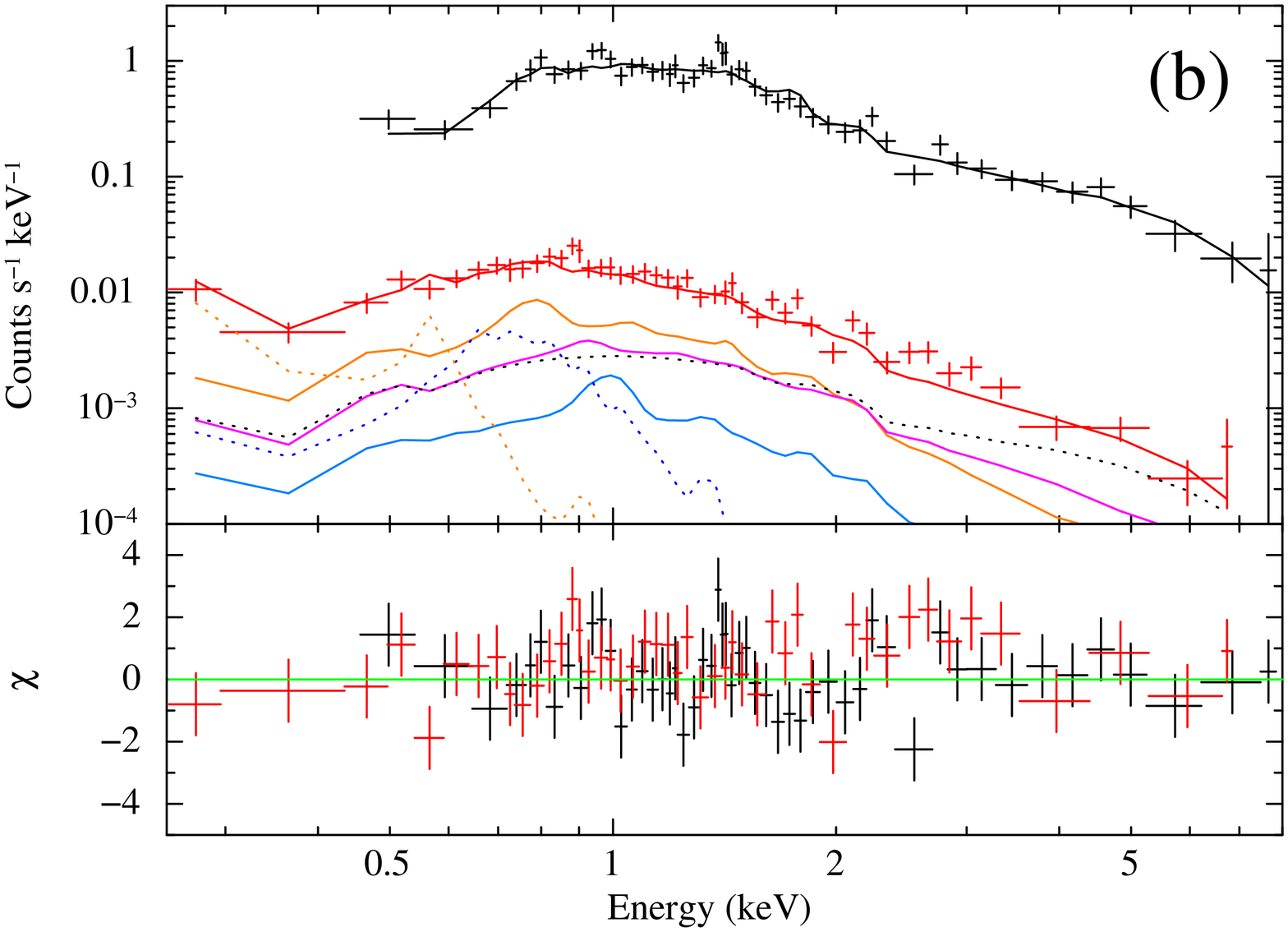}
   \end{center}
   \end{minipage}
 \caption{Spectra of the SW1 (left) and SW2 (right) region after the
 subtraction of the NXB. The FI (black) and BI (red) spectra are
 simultaneously fitted with a model of the ICM + CXB (black dashed
 line)+MWH (blue dashed line) + LHB (orange dashed line) +
 [contamination from another region]. The blue, orange and magenta solid
 lines indicate a component of emission or contamination from the SW1 region,
 that from SW2 and that from Center, respectively. The green line of the
 left panel
 represents the unfolded model of the component of emission from the SW1
 region in the arbitrary unit. The FI spectrum is multiplied by 100 for display purpose.}
 \label{spec_SW}
\end{figure}

We employ similar data reduction and analysis for the spectra 
of SW1 and SW2 regions to those of the Center region
(Section 3.1). Since the X-ray emission in each SW region is far fainter
than that in the Center region, contamination (pile-over) from the
Center region should be taken into account. 
We estimate it 
by using the best-fit spectral parameters in Table~\ref{tab:Center_result}
and the ARFs of which source region is set at the Center region. 
We also have to take the mutual contamination between SW1 and SW2 
into account. The spectra of SW1 and SW2 are thus fitted simultaneously.
Figure~\ref{spec_SW} shows the NXB-subtracted spectra of SW1 and SW2
with their best-fit models. The best-fit parameters are
summarized in Table~\ref{tab:SW_result}. In the SW1 region, the He-like Fe L-shell emissions
around 1\,keV determine the redshift  to be
$z=0.11\pm0.02$, which is consistent with $z=0.1066$
obtained from the optical observation (\cite{Struble_Rood_1999}). The
temperature and metal abundance are  $2.0\pm0.2$\,keV and 
$0.41^{+0.17}_{-0.13}$\,$Z_\odot$. 
The temperature is significantly lower than that in
the Center region. The metal abundance is marginally different from
that in the Center region but is typical for clusters. 

On the other hand, the X-ray spectrum of the SW2 region
shows no apparent line-like structures. This is partly due to its lower surface 
brightness than the SW1 region and 
also to the strong contamination from the Center region; the contamination 
from the Center region is comparable to the intrinsic emission from the
SW2 regrion. We thus conclude that the
X-ray spectral property of the SW2 region is hard to be constrained with 
this observation.

\subsection{Radial Profiles}
\label{subsec:radial}
Spatial structures of the extended X-ray emission in the Center region and
the SW1 region are examined by fitting their radial profiles with 
a $\beta$-model (\cite{Cavaliere_1978}). This procedure is also needed to 
evaluate their X-ray luminosities. 
Radial profiles were derived from the X-ray image shown in 
Figure~\ref{A1674_with_region}.
The radial profiles derived in this way are those convolved with the Suzaku PSF.
We thus estimate the Suzaku PSF with the FTOOL {\it{xissimarfgen}} and 
convolve a $\beta$-model with the PSF to fit the radial profile,
as was done by \cite{Mori_2013}. 

\begin{figure}
   \begin{minipage}{0.5\hsize}
   \begin{center}
      \FigureFile(80mm,50mm){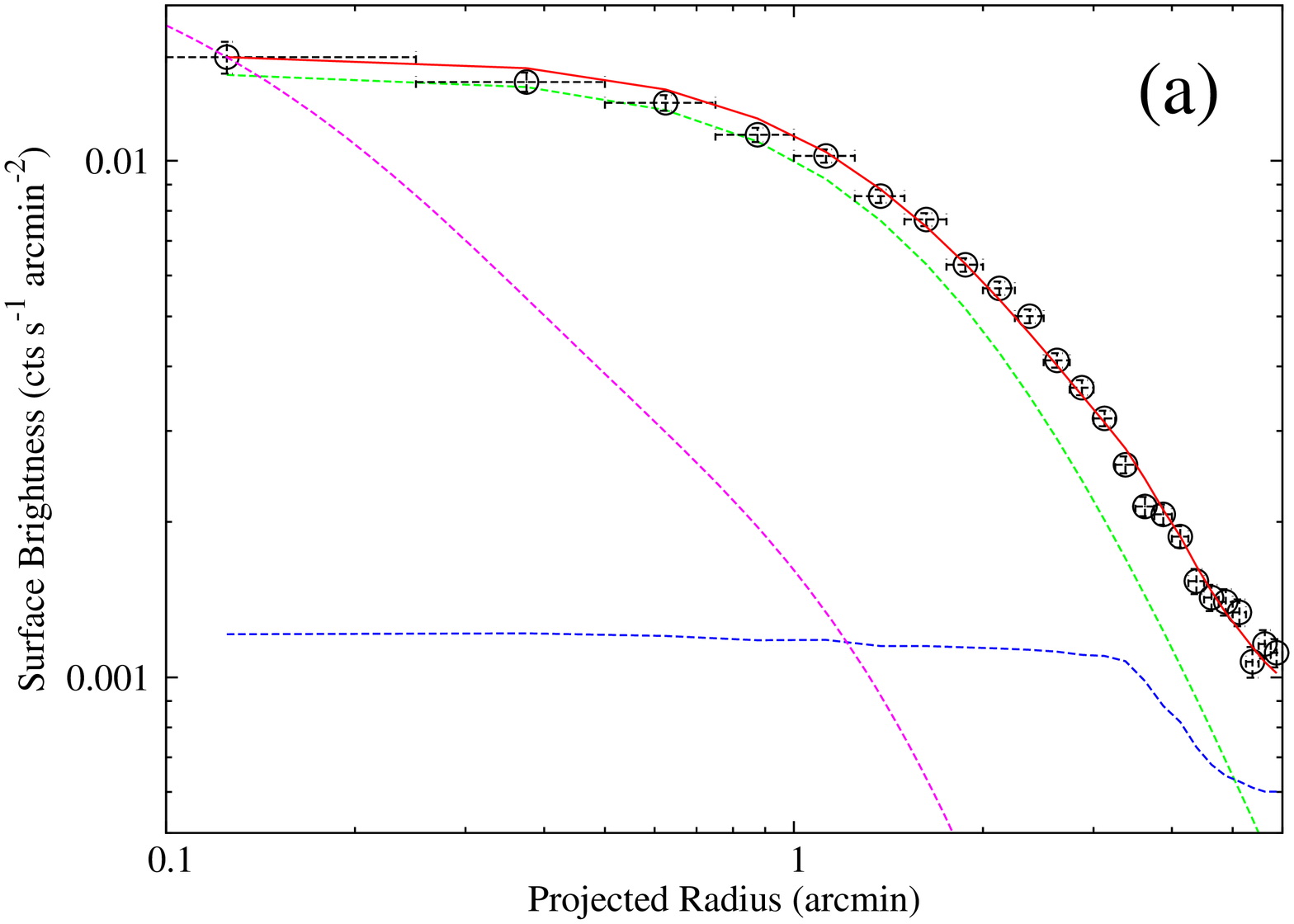}
   \end{center}
   \end{minipage}
      \begin{minipage}{0.5\hsize}
   \begin{center}
      \FigureFile(80mm,50mm){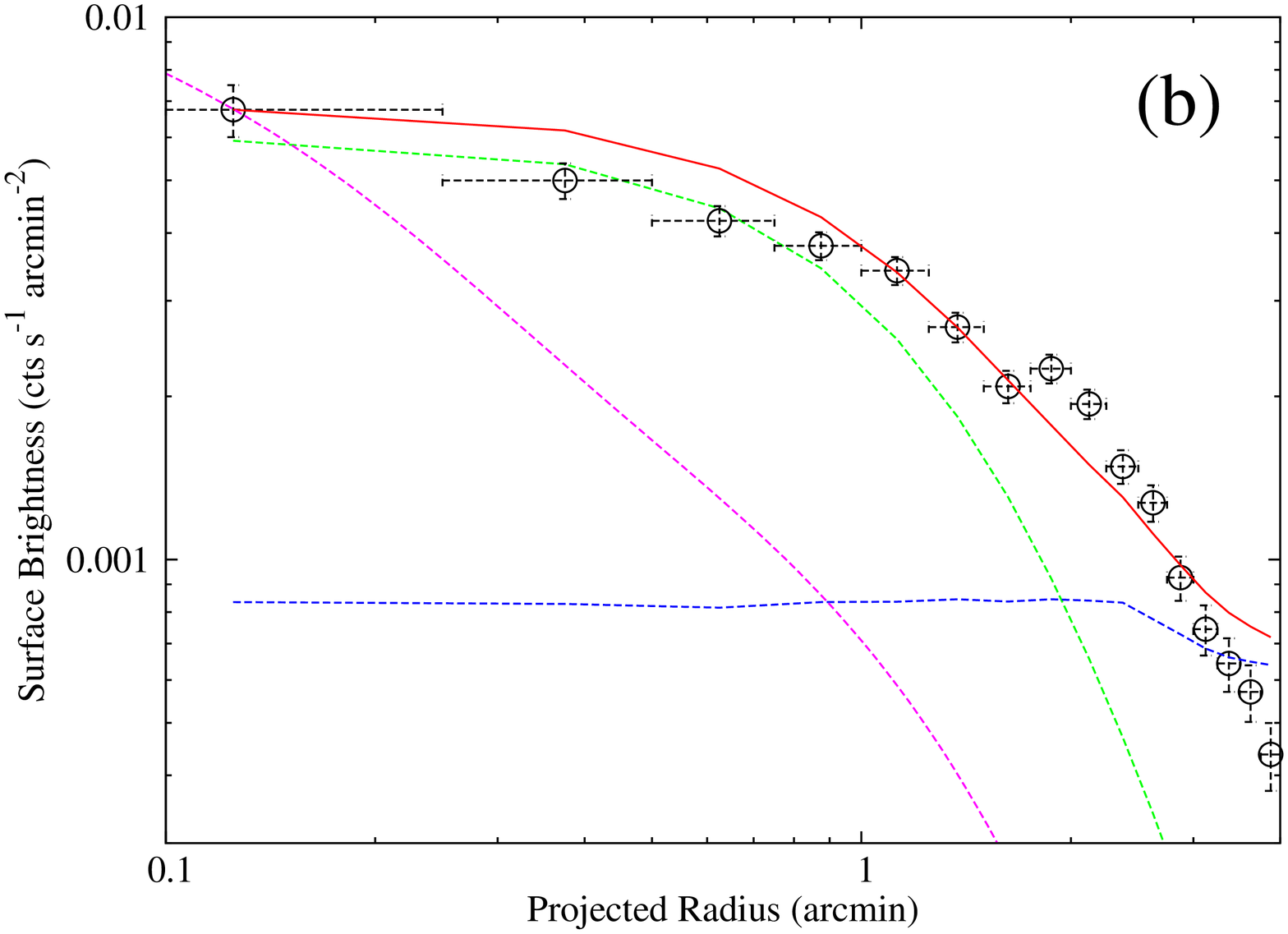}
   \end{center}
   \end{minipage}
 \caption{X-ray surface brightness in the Center (left) and SW1 (right)
 regions (open circles). The X-ray image of Fig.~\ref{A1674_with_region}
 (energy ranges are $0.4-10$\,keV for the FI-CCDs and $0.25-8$\,keV for
 the BI-CCD) is employed, and the NXB is subtracted. The error bar for each data shows a 1 $\sigma$ statistical uncertainty. 
The green and blue dash lines represent the convoleved $\beta$-model and XB component, respectively, and the red solid line shows the combined model of these components. The magenta dash line indicates the Suzaku PSF.}
 \label{radial}
\end{figure}

Figure~\ref{radial}~(a) shows the radial profile in the Center region
with the best-fit model, whose parameters are in the column 1 of
Table~\ref{tab:radial_result}. 
We confirm the emission of the Center region extends at least to 1.3\,Mpc from the center.
The core radius $r_c$ and $\beta$ is $450\pm40$\,kpc and $0.52\pm0.04$, respectively.
According to the analysis of a large sample of clusters
(e.g. \cite{Mohr_1999}, \cite{Ota_2004}, \cite{Akahori_2006}, \cite{Ota_2006}), 
$r_c$ and $\beta$ are in $20-800$\,kpc and in $0.2-1.2$, respectively.
It indicates that $r_c$ and $\beta$ of the Center region are within these ranges.
Figure~\ref{radial}~(b) shows the radial profile of A1674-SW with the best-fit model,
 whose parameters are in the column~2 of Table~\ref{tab:radial_result}. 
We find the $\beta$-model component dominates the background emission 
up to $\sim0.4$\,Mpc in the SW1 region.  
The core radius $r_c$ and $\beta$ of the SW1 region 
is $220^{+90}_{-70}$\,kpc and $0.9^{+0.4}_{-0.2}$, respectively. These values are also 
consistent with other clusters. We find some deviation of the radial profile 
in the SW1 region from the best fit $\beta$-model. The profile is better 
reproduced with a double $\beta$-model. Although we employ the single
$\beta$-model fit in the following discussion, we note that the
integrated X-ray luminosity based on it is lower than that with the
double $\beta$-model fit by about 10\%.

\section{Discussion}
As described in Sec.~\ref{sec:results}, we find two different redshifts,
$0.215^{+0.007}_{-0.006}$ and $0.11\pm0.02$, from the two
separate regions (the Center and SW1 regions) in the field of
A1674. This redshift difference corresponds to  $\sim 30,000$\,${\rm km\
s^{-1}}$, which is too large to be attributed to a bulk motion in
one cluster. We therefore can safely conclude that A1674 consists of at least 
two independent clusters overlapping along the line of sight. 
In the following discussion,  we name these clusters A1674-C and A1674-SW. 
We conventionally call A1674-SW a cluster in this paper, although we 
may had better call it  a group of galaxies, 
considering its lower temperature of $2.0\pm0.2$\,keV.  
Note also that we were not able to constrain the spectral properties of
the faint X-ray emission in the SW2 region.
 
Our hypothesis that A1674 is separated into two independent clusters of
A1674-C and A1674-SW is supported by the spatial distribution of
galaxies between $0.10<z<0.12$ and $0.21<z<0.23$ as shown in
Fig.~\ref{A1674_with_SDSS}~(b). Although the SDSS spectroscopic survey
is not deep enough to detect many galaxies with $z\sim0.2$
(\cite{Strauss_2002}), most of the galaxies measured in the SDSS
spectroscopic survey are spatially separated into two groups of
$0.21<z<0.23$ (A1674-C) and $0.10<z<0.12$ (A1674-SW). 
The separation is, however, not clear, and there are some overlaps. 
For example, we find one galaxy with $0.21<z<0.23$ in the SW1
region. The galaxy is at a distance of
1.24\,Mpc from the center of A1674-C. Assuming the distribution of galaxies
follows the X-ray radial profile we measured, the ratio of the number of 
galaxies inside and outside 1.1\,Mpc
should be about 3:2, which is statistically consistent with what we observe.  
We hence consider that this galaxy is most likely a member of A1674-C.     
On the other hand, we find one galaxy with $0.10<z<0.12$ in the Center region. 
The galaxy is at a distance of 1.26\,Mpc from the center of A1674-SW. 
Since we only confirmed the X-ray emission of A1674-SW up to $\sim0.4$\,Mpc,
the galaxy may not be a member of A1674-SW but a field galaxy.
Deeper spectroscopic observations of galaxies in the field of A1674 are needed 
for further discussions of optical properties of these clusters.

X-ray properties (temperature, core radius, and $\beta$) of A1674-C and A1674-SW
are within the range of those observed for other clusters, as described in Sec.~\ref{sec:results}.
The X-ray luminosity of each cluster is evaluated with their radial profiles described in 
Sec.~\ref{subsec:radial}.
If we extrapolate the radial profile up to $r=2$\,Mpc,
we obtain the X-ray luminosity  ($0.1-2.4$\,keV within 2\,Mpc) of A1674-C
to be $L=15.9\pm0.6\,\times 10^{43}\,{\rm erg\,s^{-1}}$, while that of A1674-SW is 
$1.25\pm0.07\,\times 10^{43}\,{\rm erg\,s^{-1}}$.
If we limit the outer radius to be 1.3\,Mpc and 0.4\,Mpc for A1674-C and A1674-SW,
respectively, the  X-ray luminosities of these two clusters are
$13.6\pm0.5\,\times 10^{43}\,{\rm erg\,s^{-1}}$ and $1.08\pm0.06\,\times 10^{43}\,{\rm erg\,s^{-1}}$,
respectively.
On the other hand,  expected luminosities within 2\,Mpc from the observed temperatures (3.8\,keV
for A1674-C and 2.0\,keV for A1674-SW) using the 
$L-T$ relation (equation (4) in \cite{Ikebe_2002}), are  
$7.9\times10^{43}\,{\rm erg\,s^{-1}}$ and 
$1.6\times10^{43}\,{\rm erg\,s^{-1}}$. These are consistent with the observed 
luminosities within a factor of 2, which is comparable to or smaller than the scatters of the data points found in \citet{Ikebe_2002}. 
This is  another support for our hypothesis. 

New clusters of galaxies were serendipitously discovered or identified
with Suzaku observations, as reported by \citet{Yamauchi_2010},
\citet{Yamauchi_2011} and \citet{Mori_2013}. In either case, an iron
emission line in the X-ray spectra and an extended X-ray emission are
the key to identify the source as a cluster. The discovery of the new cluster
of galaxies A1674-C is similar to those previous cases but from the previously cataloged cluster A1674.

\section{Summary}
We performed a spectral and spatial analysis for the cluster of galaxies A1674 in
separate regions observed with Suzaku. 
We discovered the He-like Fe K-shell line from the Center
region of this cluster for the first time, and find that the X-ray
 spectrum yields a higher redshift of $0.215^{+0.007}_{-0.006}$ than 
0.1066 that determined in the optical observations. 
On the other hand, the X-ray spectrum of the SW1 region is fitted with 
a redshift of $0.11\pm0.02$, primarily determined with the He-like Fe L-shell lines.
The galaxies in the SDSS spectroscopic survey data also shows
two separate component in the redshifts and spatial distributions.  
We hence conclude that A1674 consists of two independent clusters of galaxies, 
A1674-C and A1674-SW, overlapping along the line of sight.
The gas temperature, metal abundance, core radius, $\beta$ of A1674-C 
are $3.8\pm0.2$\,keV, $0.20\pm0.05$\,$Z_\odot$, 
$450\pm40$\,kpc, and $0.52\pm0.04$, respectively. 
Those of A1674-SW are $2.0\pm0.2$\,keV, $0.41^{+0.17}_{-0.13}$\,$Z_\odot$, 
$220^{+90}_{-70}$\,kpc, and  $0.9^{+0.4}_{-0.2}$, respectively.
These parameters are within the range of those observed in other clusters.
The X-ray luminosities evaluated for A1674-C and A1674-SW
are consistent with those expected from their temperatures by using 
the $L-T$ relation described in \citet{Ikebe_2002} within a factor of 2.
These are additional supports for our conclusion 
that A1674 consists of two independent clusters of galaxies.

Previous optical determination of the redshift of A1674 based on
measurements of
two member galaxies by \citet{Huchra_1990} turned out to be not 
enough for this kind of overlapping system.  X-ray observations with
high signal to noise ratio are efficient
to find different redshift components in clusters.

\section*{Acknowledgments}

We thank all members of the Suzaku operation and calibration teams.
This work is supported by Japan Society for the Promotion of Science
(JSPS) KAKENHI Grant Number 23340071 (Kiyoshi Hayashida), 12J01190
(Shutaro Ueda), 23000004 (Hiroshi Tsunemi), 24540229 (Katsuji
Koyama). SRON is also supported financially by NWO,
the Netherlands Organization for Scientific Research.
H.A. is supported by a Grant-in-Aid for Japan Society for the Promotion
of Science (JSPS) Fellows (26-606).

\newpage

\begin{table}
  \begin{center}
  \caption{Fitting results of the Center region}\label{tab:Center_result}
    \begin{tabular}{lll}
      \hline
      Component (Model) & Parameter & Value\footnotemark[$*$] \\
      \hline \hline
      ICM ({\it{apec}}) & $kT$ (keV) & $3.8\pm0.2$ \\
      & Abundance (solar) & $0.20\pm0.05$\\
      & Redshift & $0.215^{+0.007}_{-0.006}$\\
      & Luminosity\footnotemark[$\dagger$] & $15.9\pm0.6$\\
      \hline
      CXB ({\it{powerlaw}}) & Photon index & $1.412$ (fix) \\
      & Unabsorbed flux\footnotemark[$\ddagger$] & $1.94\times10^{-11}$ (fix) \\
      \hline
      MWH ({\it{apec}}) & $kT$ (keV) & $0.29^{+0.07}_{-0.05}$ \\
      & Abundance (solar) & $1.0$ (fix)\\
      & Redshift & $0.0$ (fix)\\
      & Unabsorbed flux\footnotemark[$\ddagger$] & $6.8^{+3.1}_{-2.5}\times10^{-15}$\\
      \hline
      LHB ({\it{apec}}) & $kT$ (keV) & $0.10^{+0.03}_{-0.02}$ \\
      & Abundance (solar) & $1.0$ (fix) \\
      & Redshift & $0.0$ (fix) \\
      & Unabsorbed flux\footnotemark[$\ddagger$] & $4.6^{+5.2}_{-2.2}\times10^{-19}$\\
      \hline
      Absorption ({\it{wabs}}) & $N_{\rm H}$ ($\times 10^{20}\ {\rm cm^{-2}}$) & $1.7$ (fix) \\
      \hline
      $\chi^2$/d.o.f & & 740.25/737 \\
      \hline
   \multicolumn{3}{@{}l@{}}{\hbox to 0pt{\parbox{85mm}{\footnotesize
      \vspace{3mm}
      \par\noindent
      \footnotemark[$*$]: The errors represent the 90\% confidence
     range.
      \par\noindent
      \footnotemark[$\dagger$]: The luminosity is represented in a unit of
     $10^{43}\,{\rm erg}\,{\rm
     s}^{-1}$ in the $0.1-2.4$\,keV band within $r=2$\,Mpc, where the radial profile is taken into account. 
      \par\noindent
      \footnotemark[$\ddagger$]: Flux $({\rm erg}\,{\rm
     cm}^{-2}\,{\rm
     s}^{-1}\,{\rm deg}^{-2})$ are in the $2-10$\,keV band.
    }\hss}}
    \end{tabular}
  \end{center}
\end{table}

\begin{table}
  \begin{center}
  \caption{Results of the simultaneous fit of the SW1 and SW2 region.}\label{tab:SW_result}
    \begin{tabular}{llll}
      \hline
      Parameter & SW1\footnotemark[$*$] & SW2\footnotemark[$*$]\\
      \hline \hline
      $kT$ (keV) & $2.0\pm0.2$ & $2.1^{+0.4}_{-0.6}$\\
      Abundance (solar) & $0.41^{+0.17}_{-0.13}$ & $0.26^{+0.21}_{-0.19}$\\
      Redshift & $0.11\pm0.02$ & $0.41^{+0.02}_{-0.15}$\\
      Luminosity\footnotemark[$\dagger$] & $1.25\pm0.07$ & \multicolumn{1}{c}{-----}\\
      \hline
      $\chi^2$/d.o.f & \multicolumn{2}{c}{456.6/429} \\
      \hline
   \multicolumn{3}{@{}l@{}}{\hbox to 0pt{\parbox{85mm}{\footnotesize
      \vspace{3mm}
      \par\noindent
      \footnotemark[$*$]: The errors represent the 90\% confidence
     range.
      \par\noindent
      \footnotemark[$\dagger$]: The luminosity is represented in a unit of
     $10^{43}\,{\rm erg}\,{\rm
     s}^{-1}$ in the $0.1-2.4$\,keV band within $r=2$\,Mpc, where the radial profile is taken into account. 
    }\hss}}
    \end{tabular}
  \end{center}
\end{table}

\begin{table}
  \begin{center}
  \caption{Radial profiles of the SW1 and SW2 region fitted with a $\beta$
   model.}\label{tab:radial_result}
    \begin{tabular}{lll}
      \hline
      Parameter & Center\footnotemark[$*$] & SW1\footnotemark[$*$]\\
      \hline \hline
      Core radius $r_c$ (kpc) & $450\pm40$ & $220^{+90}_{-70}$\\
      $\beta$-index & $0.52\pm0.04$ & $0.9^{+0.4}_{-0.2}$\\
      \hline
      $\chi^2$/d.o.f & 46.3/22 & 95.5/14\\
      \hline
   \multicolumn{3}{@{}l@{}}{\hbox to 0pt{\parbox{85mm}{\footnotesize
      \vspace{3mm}
      \par\noindent
      \footnotemark[$*$]: The errors represent the 90\% confidence
     range.      
    }\hss}}
    \end{tabular}
  \end{center}
\end{table}

\end{document}